\documentclass[sigconf,screen,authorversion]{acmart}
\usepackage{subcaption}

\AtBeginDocument{%
  \providecommand\BibTeX{{%
    \normalfont B\kern-0.5em{\scshape i\kern-0.25em b}\kern-0.8em\TeX}}}

\copyrightyear{2023}
\acmYear{2023}
\setcopyright{acmlicensed}

\acmConference[ETRA '23]{2023 Symposium on Eye Tracking Research and Applications}{May 30-June 2, 2023}{Tubingen, Germany}

\acmBooktitle{2023 Symposium on Eye Tracking Research and Applications (ETRA '23), May 30-June 2, 2023, Tubingen, Germany}
\acmPrice{15.00}
\acmDOI{10.1145/3588015.3589511}
\acmISBN{979-8-4007-0150-4/23/05}

\begin{document}

\title[Eye Tracking as a Source of Implicit Feedback in Recommender Systems]{Eye Tracking as a Source of Implicit Feedback in Recommender Systems: A Preliminary Analysis}

\author{Santiago de Leon-Martinez}
\affiliation{%
  \institution{Faculty of Information Technology, Brno University of Technology}
  \city{Brno}
  \country{Czechia}
}
\additionalaffiliation{%
  \institution{Kempelen Institute of Intelligent Technologies}
  \city{Bratislava}
  \country{Slovakia}
}
\email{santiago.deleon@kinit.sk}
\orcid{0000-0002-2109-9420}

\author{Robert Moro}
\affiliation{%
  \institution{Kempelen Institute of Intelligent Technologies}
  \city{Bratislava}
  \country{Slovakia}
}
\email{robert.moro@kinit.sk}
\orcid{0000-0002-3052-8290}

\author{Maria Bielikova}
\affiliation{%
  \institution{Kempelen Institute of Intelligent Technologies}
  \city{Bratislava}
  \country{Slovakia}
}
\email{maria.bielikova@kinit.sk}
\orcid{0000-0003-4105-3494}

\renewcommand{\shortauthors}{de Leon-Martinez et al.}

\begin{abstract}
Eye tracking in recommender systems can provide an additional source of implicit feedback, while helping to evaluate other sources of feedback. In this study, we use eye tracking data to inform a collaborative filtering model for movie recommendation providing an improvement over the click-based implementations and additionally analyze the area of interest (AOI) duration as related to the known information of click data and movies seen previously, showing AOI information consistently coincides with these items of interest.
\end{abstract}



\maketitle

\section{Introduction \& Related Works}
An ever present problem in recommender systems (RS) is the evaluation of implicit and explicit feedback and assumptions used when processing user feedback. For example, the popular cascade click model assumes that users have seen (or skipped) every item in the list before the clicked one and none after~\cite{richardson_predicting_2007,craswell_experimental_2008}; however, this assumption can easily be verified using eye tracking (ET). In web retrieval, ET was fundamental to showing correspondence between clicks and explicit judgements~\cite{joachims_accurately_2016,joachims_evaluating_2007}. In RS, studies using ET have been focused on analyzing the user behavior within a RS interface~\cite{castagnos_eye-tracking_2010,guan_eye_2007}, inferring user traits~\cite{chen_eye-tracking-based_2022, millecamp_classifeye_2021}, or predicting the users' gaze or interest~\cite{zhao_gaze_2016,li_towards_2017}. However, the area is still under-developed especially in regards to better interpreting users' implicit feedback and part of this is due to a lack of public recommendation datasets available with ET data. 

\begin{figure*}[!tbp]
  \begin{subfigure}[b]{\columnwidth}
    \centering
    \includegraphics[width=0.56\textwidth]{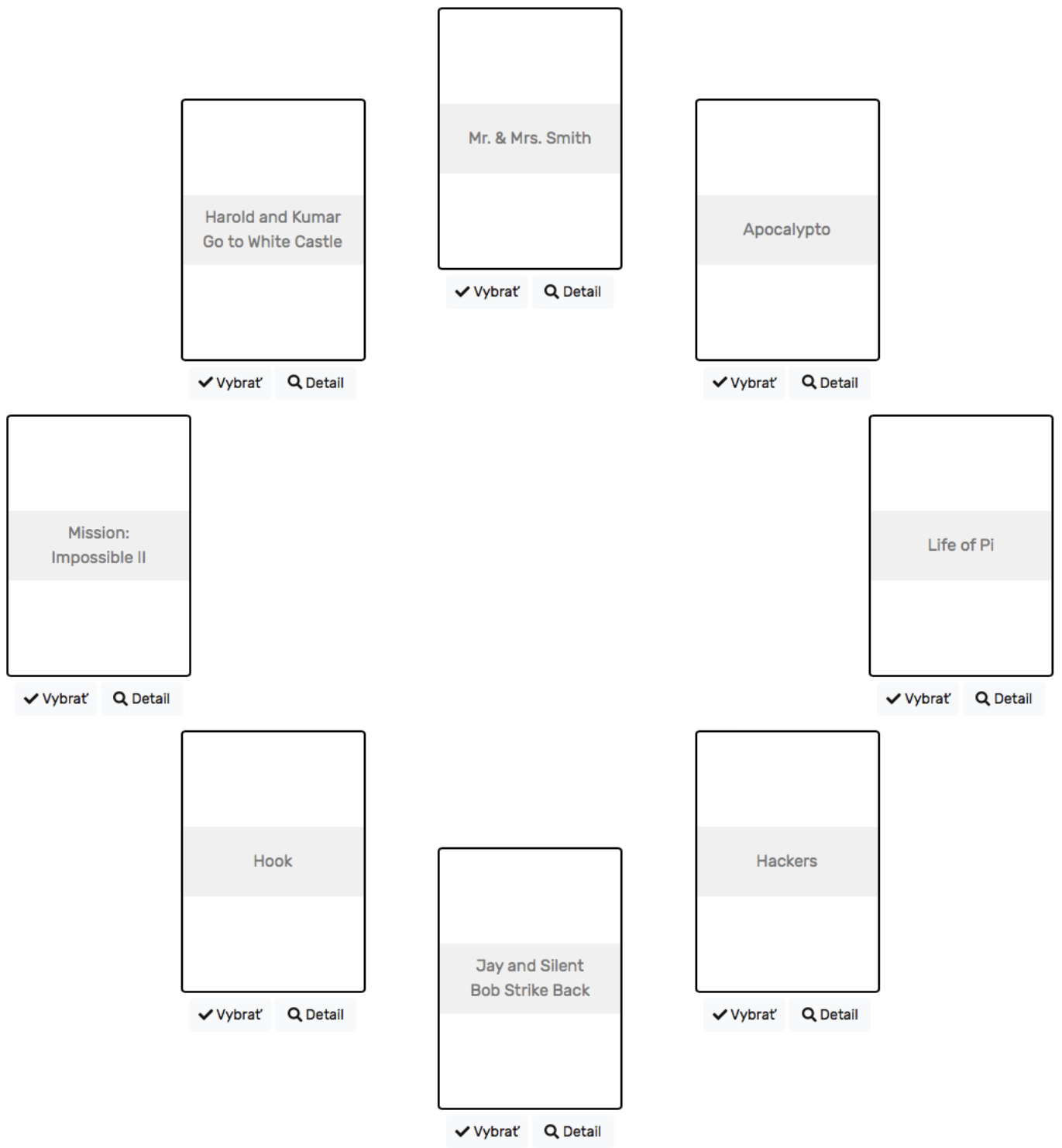}
    \label{fig:f1}
  \end{subfigure}
  \begin{subfigure}[b]{\columnwidth}
    \centering
    \includegraphics[width=0.45\textwidth]{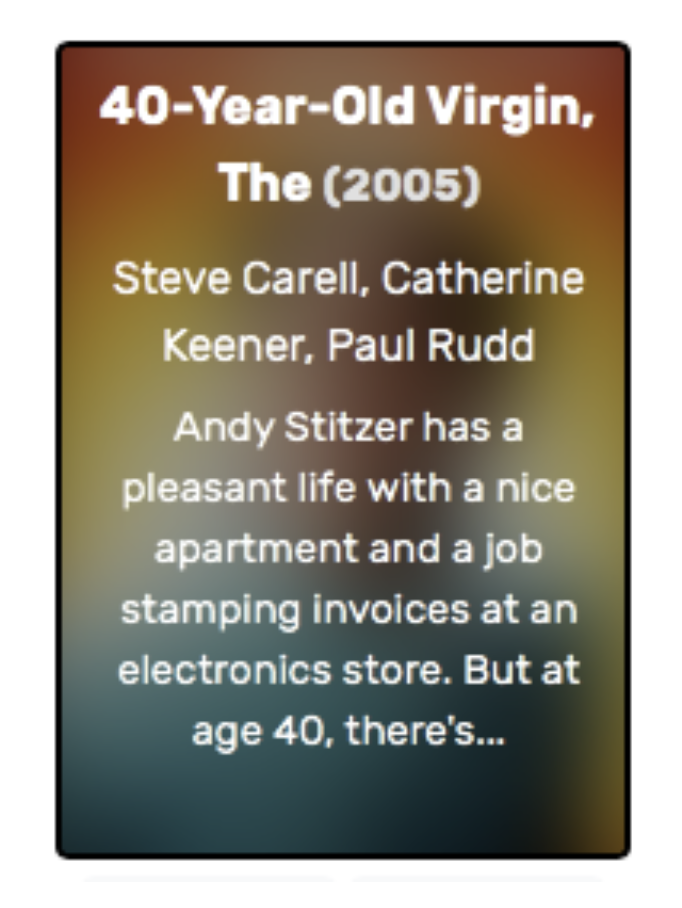}
    \label{fig:f2}
  \end{subfigure}
  \caption{An example of the circular movie list with buttons for selection and detail showing the setup with textual titles (left) \& an example of the details displayed (right).}
  \label{fig:big}
\end{figure*}

In this work, we aim to showcase the potential of using ET data as a source of implicit feedback for RS. We build upon an existing study of ET data within a RS that examined gaze patterns and positional bias in circular movie lists of text only and images, as seen in Figure~\ref{fig:big} \cite{gaspar_analysis_2018}. Our contribution is the application of this ET data for generating recommendations, while the previous study was only observational with no such application.

\section{Experimental Methods}
The study in~\cite{gaspar_analysis_2018} employed a 2x2 within-subjects design. It asked users to select a movie from a randomly sorted circular list; the movie lists were generated randomly or from a user's preferred categories of movies. The users were presented with poster images (12 screens) and textual titles (12 screens) with the ability to click for movie details (see Figure~\ref{fig:big}). In total, there were 64 participants (45 males, 19 females). ET data was collected using the Tobii X2-60 60Hz eye trackers mounted upon screens with the resolution of 1920x1200px.

We used the same methods for processing the gaze data for fixations translated to movie areas of interest (AOIs) and also used the gathered click data including the final movie selected and movies clicked for details. Information of movies already seen by users was also provided. After filtering the dataset of errors in gaze data and screens where the user did not pick a movie, 55 users and 1159 screens remained. Total duration spent at each AOI was calculated by screen; the mean $\mu$ and standard deviation $\sigma$ were calculated for a user across screens of image and text separately.

In the experiment, we aimed to rank the movies presented on the screen based on the users' interests inferred from their previous movie interactions gathered from other screens. For this purpose, we used a collaborative filtering (CF) model using matrix factorization~\cite{koren_matrix_2009} with bias. We hypothesized that the additional movie interactions learned through gaze data, where fixation time may be related to attention/interest, would lead to better recommendation lists with the selected movie earlier in the ranking. To evaluate this hypothesis, we compared the performance of several interaction filtering methods based on the movie AOIs' fixation duration. They differed in what movies were included for a given user in the training set by including movies that were fixated more than a duration threshold $\tau$ with $\tau$ equal to $\mu+\sigma$, $\mu$, and $\mu-\sigma$ respectively. We evaluated them against the non-AOI baselines using click information only.

MovieLens 20M~\cite{harper_movielens_2016}, a 138,000 users by 27,000 movies dataset containing 20 million ratings, was used as a training set for the model as it included all movies in the study, but to reduce computational burden it was first filtered  to users that had ratings 4.0 or greater of the movies included in the study, which we then binarized (a common practice in movie recommendation, see, e.g.,~\cite{ferrari_dacrema_troubling_2021,liang_variational_2018}). A hyperparameter search was done on this filtered dataset (not including the study data) with a .8/.1/.1 split optimizing normalized discounted cumulative gain for the top 100 items (NDCG@100). 
For the experiment, the study data was joined with the filtered dataset to generate a training set. Then for each user first the interactions from the image screens were held-out and then the experiment was repeated holding-out the text screens' interactions. This led to a total of 110 models being trained per interaction filtering method, which were then used to rank the movies of the held-out image or text screens. Evaluation was done by comparing ranking metrics averaged across test screens of the movie selected (relative to the other 7 movies) between the different interaction filtering methods.

\section{Results \& Discussion}
In Table \ref{tab:results}, results of the CF experiment with AOIs information based on the three different filtering methods are compared to the non-AOI baselines using only click data. The least restrictive AOI threshold by user mean total duration minus one standard deviation performed best on Mean Recall@1,2,4 and achieved the lowest average ranking position (of values 1 to 8), while the threshold filter of just mean performed the worst even compared to the random baseline.

\begin{table*}[t]
\caption{Results of CF with matrix factorization on a test set comparing the rank of the movie selected to other movies presented in the same screen using different methods of implicit feedback.}
\label{tab:results}
\resizebox{\linewidth}{!}{%
\begin{tabular}{lccccc}
\hline
 &
  \multicolumn{1}{l}{Mean Recall@1 (Std)} &
  \multicolumn{1}{l}{Mean Recall@2 (Std)} &
  \multicolumn{1}{l}{Mean Recall@3 (Std)} &
  \multicolumn{1}{l}{Mean Recall@4 (Std)} &
  \multicolumn{1}{l}{Mean Rank (Std)} \\ \hline
Baseline (Random)                  & 12.50 & 25.00 & 37.50          & 50.00 & 4.500  \\
Baseline (Selected)                           & 13.04 (0.35) & 24.44 (0.45) & 38.08 (0.49)          & 50.17 (0.50) & 4.489 (0.13) \\
Baseline (Selected, Detailed)                 & 13.2 (0.33) & 26.06 (0.42) & \textbf{38.48 (0.48)} & 49.7 (0.50) & 4.488 (0.13) \\
Selected, Detailed, AOIs $\mu+\sigma$ & 13.03 (0.33) & 24.94 (0.43) & 37.62 (0.48)          & 49.44 (0.50) & 4.494 (0.13) \\
Selected, Detailed, AOIs $\mu$      & 12.08 (0.33) & 25.28 (0.43) & 35.98 (0.49)         & 49.27 (0.50) & 4.541 (0.13)  \\
Selected, Detailed, AOIs $\mu-\sigma$ &
  \textbf{14.06 (0.34)} &
  \textbf{27.35 (0.43)} &
  37.96 (0.48) &
  \textbf{51.68 (0.50)} &
  \textbf{4.415 (0.13)} \\ \hline
\end{tabular}%
}
\end{table*}

Additionally, we analyzed the number of movies included by the AOI filtering methods to determine the information provided by each. Inclusion percentages of movies that had been selected, clicked for details, and previously seen were calculated per screen then averaged across all screens. We additionally included information on the AOIs without filtering and all movies presented in the selection. The unfiltered AOIs cover 96.05\% of movies in the list showing users were at least briefly fixating on almost all movies. In regards to the filters, they were successful in selecting for informative AOIs. In particular, the selected ($\mu+\sigma$: 50.65\%, $\mu$: 79.12\%, $\mu-\sigma$: 98.27\%) and detailed ($\mu+\sigma$: 54.73\%, $\mu$: 83.49\%, $\mu-\sigma$: 97.7\%) movies were comparatively retained more than all movies ($\mu+\sigma$: 13.43\%, $\mu$: 31.13\%, $\mu-\sigma$: 93.38\%). This is to be expected as movies that were selected and detailed would be those that attracted the attention of the user and, as we hypothesized, it would also be more likely that the user spends time examining the movie information. Examining seen movies, it appears that the increase is comparatively less than the selected and detailed movies ($\mu+\sigma$: 15.15\%, $\mu$: 37.38\%, $\mu-\sigma$: 94.04\%), but is still a 20\% increase from the all movies in both the $\mu+\sigma$ and $\mu$ filters. We postulate that seen movies may draw some attention from the user (as they are more likely to be included in the fixation thresholds than movies without distinction), but do not hold the attention of the user like the selected or detailed movies.

Furthermore, when comparing the results of this analysis with the CF experiment, we propose that the least restrictive filter is most beneficial to the binarized CF model due to the higher inclusion of seen movies in training. Selection of a movie is positive feedback, but it does not directly imply consumption, while having seen a movie and rating a movie do. As mentioned before, ET in recommender systems provides beneficial data from which to evaluate which items were processed by the user. We argue that time spent at AOIs may correspond with attention and interest in a movie that later leads to consumption, but it also provides the benefit of excluding movies that were not fixated or briefly fixated. 

\section{Conclusion}
In this study, we used ET data based on movie AOI durations as an additional source of implicit feedback to enrich a CF model providing better recommendations more representative of the movies selected within the study. We further analyzed the AOI durations to find that it contains relevant information for recommendation across different filtering techniques. In terms of the future, it would be beneficial to gather more ET data within a RS setting, as it is currently dwarfed by common datasets in RS. Additionally we would like to implement a probabilistic click model taking into account the ET feedback and use this to validate and learn model aspects, such as skipping, positional bias, and more.

\begin{acks}
This work was supported by Eyes4ICU, a project funded by the European Union under the Horizon Europe Marie Sk\l{}odowska-Curie Actions, GA No. \href{https://doi.org/10.3030/101072410}{101072410}. We would also like to acknowledge Peter Gaspar for his previous work, willingness to share the dataset that this study used, and help therein.
\end{acks}

\bibliographystyle{ACM-Reference-Format}
\bibliography{references}

\end{document}